\documentclass[conference]{IEEEtran}
\IEEEoverridecommandlockouts
\usepackage{algorithmic}
\usepackage{amsmath,amssymb,amsfonts}
\usepackage{cite}
\usepackage{graphicx}
\usepackage{multirow}
\usepackage{textcomp}
\usepackage{xcolor}

\newcommand\blfootnote[1]{
  \begingroup
  \renewcommand\thefootnote{}\footnote{#1}
  \addtocounter{footnote}{-1}
  \endgroup
}

\def\BibTeX{{\rm B\kern-.05em{\sc i\kern-.025em b}\kern-.08em
    T\kern-.1667em\lower.7ex\hbox{E}\kern-.125emX}}

\begin{document}

\makeatletter
\newcommand{\linebreakand}{
    \end{@IEEEauthorhalign}
    \hfill\mbox{}\par
    \mbox{}\hfill\begin{@IEEEauthorhalign}
}
\makeatother

\title{The efficiency of deep learning algorithms for detecting anatomical reference points on radiological images of the head profile}

\author{
    \IEEEauthorblockN{Konstantin Dobratulin}
    \IEEEauthorblockA{
        Faculty of Informatics \\
        Samara National Research University \\
        Samara, Russia \\
        dobratulin@protonmail.com
    }
    \and
    \IEEEauthorblockN{Andrey Gaidel}
    \IEEEauthorblockA{
        Video Mining Laboratory \\
        IPSI RAS \\
        Samara, Russia \\
        andrey.gaidel@gmail.com
    }
    \and
    \IEEEauthorblockN{Aleksandr Kapishnikov}
    \IEEEauthorblockA{
        Department of Radiology and Radiotherapy \\
        with the course of Medicine Informatics \\
        FSBEI HE SamSMU MOH Russia \\
        Samara, Russia \\
        a.kapishnikov@gmail.com
    }
    \and
    \IEEEauthorblockN{Anna Ivleva}
    \IEEEauthorblockA{
        Laboratory of Analysis and Modeling \\
        of Complex Systems \\
        ICCS RAS \\
        Samara, Russia \\
        annushka199@bk.com
    }
    \and
    \IEEEauthorblockN{Irina Aupova}
    \IEEEauthorblockA{
        Department of Pediatric Dentistry \\
        FSBEI HE SamSMU MOH Russia \\
        Samara, Russia \\
        aupovaio@mail.ru
    }
    \and
    \IEEEauthorblockN{Pavel Zelter}
    \IEEEauthorblockA{
        Department of Radiology and Radiotherapy \\
        with the course of Medicine Informatics \\
        FSBEI HE SamSMU MOH Russia \\
        Samara, Russia \\
        pzelter@mail.ru
    }
}

\maketitle

\begin{abstract}
    In this article we investigate the efficiency of deep learning
    algorithms in solving the task of detecting anatomical reference points
    on radiological images of the head in lateral projection using a fully
    convolutional neural network and a fully convolutional neural network
    with an extended architecture for biomedical image segmentation -
    U-Net. A comparison is made for the results of detection anatomical
    reference points for each of the selected neural network architectures
    and their comparison with the results obtained when orthodontists
    detected anatomical reference points. Based on the obtained results, it
    was concluded that a U-Net neural network allows performing the
    detection of anatomical reference points more accurately than a fully
    convolutional neural network. The results of the detection of
    anatomical reference points by the U-Net neural network are closer to
    the average results of the detection of reference points by a group of
    orthodontists.
\end{abstract}


\blfootnote{
    The work was partially funded by the Russian Foundation for Basic Research under grants No. 18-07-01390, 19-29-01235 and 19-29-01135 (theoretical results) and the
    RF Ministry of Science and Higher Education within the government project of the FSRC “Crystallography and Photonics” RAS under grant No. 007-GZ/Ch3363/26 (numerical calculations).
}

\section{Introduction} \label{Introduction}

Teleradiography of the lateral projection of the head is the main and most informative research method used in the planning process of the orthodontic treatment of patients \cite{c01}. This method allows you to detect anomalies of occlusion in the sagittal and vertical directions, estimate the size of the upper and lower jaws, measure the length of the branches of the lower jaw, calculate the angle of inclination of the incisors of both jaws, both to the plane of the corresponding jaw and to the base of the skull, determine type of growth of the facial skeleton \cite{c01,c02,c03}. Cephalometric analysis of teleradiography of the lateral projection of the head in the lateral projection is a labor-intensive diagnostic method that requires the doctor's outstanding experience and high qualification. In the process of cephalometric analysis, the complexity of deciphering a picture arises due to the individual structural features of the facial skeleton \cite{c05}, the overlapping of anatomical structures, the presence of tooth rudiments. Improving the technologies of methods of radiology diagnostics and interpretation of the results obtained using such methods requires the introduction of modern information technologies into the practice of an orthodontist.
To simplify the process of deciphering two-dimensional skull images in a lateral projection, we propose to automate the search for supporting anatomical reference points necessary for further calculation of numerical parameters - distances and angles between reference points, which is necessary to determine the anomaly of the dentofacial system and to draw up a treatment plan for the patient.

In this work, we study methods for finding anatomical reference points on a radiological image of the skull in lateral projection using a fully convolutional neural network \cite{c06,c07}, as well as using a fully convolutional neural network with an extended architecture for segmenting biomedical images called U-Net \cite{c09}. The results of localization of anatomical reference points for each of the selected architectures of neural networks presented and their comparison with the results obtained by localizing anatomical reference points by orthodontists.

\section{Data preparation} \label{Data preparation}

\subsection{Extracting the coordinates of the anatomical reference}

The process of determining the anatomical reference points by a doctor on two-dimensional images of the skull in a lateral projection, obtained by the method of radiology \cite{c04}, is reduced to setting reference points for each type of reference point on each of the images. When processing the image, the coordinates of the point set by the doctor extracted. The coordinates of the anatomical reference points on it correspond to each initial two-dimensional image of the skull in the lateral projection. This approach allows us to apply machine learning methods further to localize anatomical reference points and to compare the results of the applied algorithms with the results obtained by the doctor.

The extraction of coordinates for each of the anatomical reference points consists of setting a point on a two-dimensional image by a doctor in a place corresponding to the anatomical reference points. In contrast, the size of the setpoint is a pixel to reduce the likelihood of erroneous localization, and then extracting the coordinates of the point.

An example of radiological image of the head in lateral projection presented in Fig.~\ref{image-example-radiological}.

\begin{figure}[htbp]
    \centerline{\includegraphics[scale=0.40]{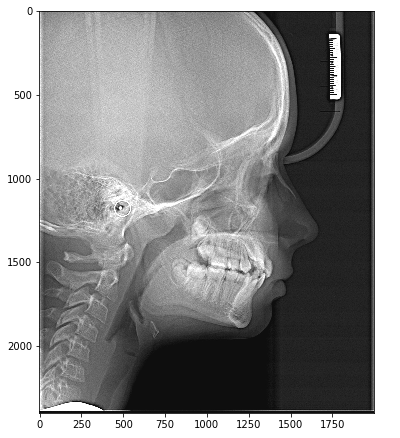}}
    \caption{Radiological image of the head in lateral projection.}
    \label{image-example-radiological}
\end{figure}

At the stage of placement of points, it is necessary to determine 27 unique positions. From the task of data sampling, the doctor needs to mark each of the points on the radiological image. Thus, 27 points differing from each other, both in coordinates and in the criteria for choosing the area of their location, fall on one radiological image \cite{c11}.
Further process of training convolutional neural networks for the task of localizing anatomical reference points requires a certain number of training samples. In this study, for the training and verification of the localization of the region of points by the model, 100 radiological images of the lateral projection of the head were selected for 100 unique patients, respectively. On each of the images, the doctor marked 27 points, after which, for each anatomical reference point, its coordinate in the image extracted. As a result, coordinates obtained for each of the 27 types of reference points in each of the 100 images.
To obtain additional statistics, three doctors with a work experience of 10 to 15 years participated in the study. Each of the doctors, regardless of the others, performed the process of marking the anatomical reference points on the proposed 100 images. In the end, three variants of setting each of the points in each picture obtained. This approach evaluating the mathematical parameters for the results of the localization of anatomical reference points by doctors.

\subsection{Dataset formation}

In this work, the marking of anatomical reference points carried out for radiological images of \(2000 \times 2400\) pixels in size, which is convenient for marking points by an orthodontist. However, it is quite resource-intensive when using images of such sizes in the process of training neural networks, as well as a further prediction on new images. To reduce computational costs, the size of the original images was changed to \(432 \times 512\) pixels using the bilinear interpolation method \cite{c08}.

For the architecture of a neural network based on the use of convolutional layers \cite{c06}, there is a high probability of losing a single pixel of an anatomical reference point when passing the convolution filter, which makes it impossible to minimize the error function and the neural network never learn to the degree necessary for prediction.
When converting the coordinates of an anatomical reference point into a two-dimensional matrix, it proposed to use not just setting a pixel of a unit value, but setting a square matrix with normally distributed values in the coordinate region. In contrast, the maximum value of the matrix reached the point with the coordinates of the anatomical reference point.
A uniform increase in the values to the real position of the anatomical reference point in a two-dimensional radiological image is obtained, which allows the use of convolutional layers in neural network models to train the localization of the anatomical reference point, and convolution can perform with a kernel of arbitrary size without losing a pixel at the position of the anatomical reference point. The advantage of this approach is the uniform increase in the values to the position of the point, that is, the absence of sharp changes in values.

An example of the formation of a matrix for a radiological image scaled to the size of the input image of a neural network model containing a portion with normally distributed values in the region of the point and reaching a maximum at a point with the coordinates of the anatomical reference point shown in Fig.~\ref{image-example-mask}.

\begin{figure}[htbp]
    \centerline{\includegraphics[scale=0.75]{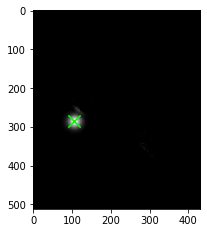}}
    \caption{Mask with normally distributed values aroung single pixel.}
    \label{image-example-mask}
\end{figure}

\section{Neural networks training} \label{Neural networks training}
After scaling the images and transforming the coordinates of the anatomical reference points into a matrix representation with a plot consisting of normally distributed values in the point region, it becomes possible to form train and test datasets for neural networks.
To effectively train neural networks that localize 27 points of anatomical reference points, we construct the architecture of neural networks as follows: the input is batches of radiological images in grayscale, size \(432 \times 512 \times 1\), and the output is batches of 27 localization masks of size \(432 \times 512 \times 27\). This approach allows the use of selected neural networks to localize 27 anatomical reference points at once.
Model training and localization results were obtained based on the cross-validation method. For the training sample, 80 out of 100 used images allocated, and for the test sample, the remaining 20 images, which in the percentage ratio exactly corresponds to the ratio 80/20. For cross-validation, 5 partitions of the initial data used; training and validation performed 5 times. Images data did not shuffled between partitions.
The neural networks has trained for 80 epochs. As an optimizer, Adam used with a learning rate of 0.001 \cite{c10}.
The data partitioning scheme for cross-validation shown in Fig.~\ref{image-split}.

\begin{figure}[htbp]
    \centerline{\includegraphics[scale=0.20]{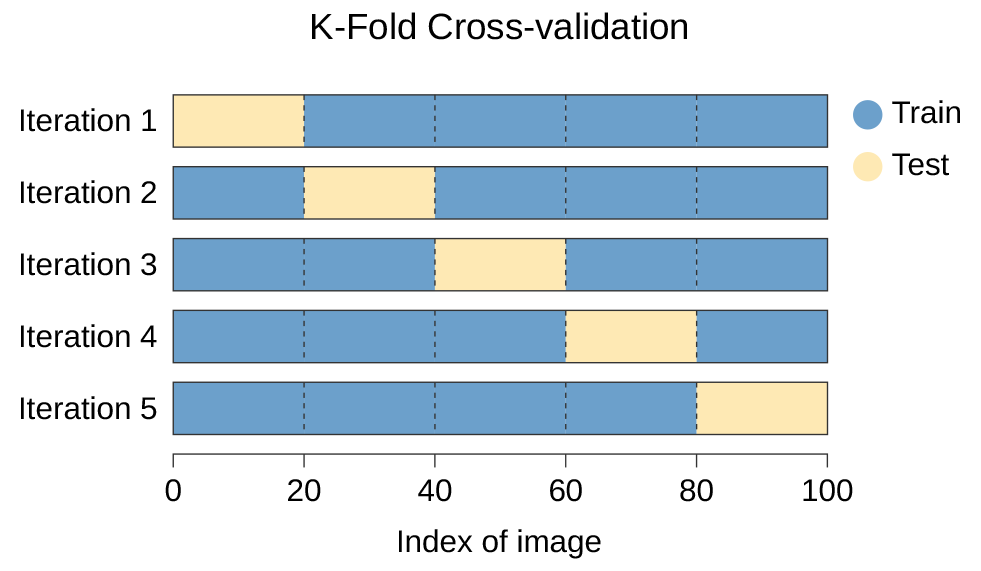}}
    \caption{Cross-validation data split scheme.}
    \label{image-split}
\end{figure}

Considered models of neural networks are conceptually similar in their structure. In this study, evaluating the effectiveness of the U-Net model compared to a fully convolutional neural network plays an important role. In the tasks of medical segmentation and localization of images, the struggle for accuracy is of great importance.
The used architecture for a fully convolutional neural network shown in Fig.~\ref{image-model-cnn}.
Train history graph for a fully convolutional neural network shown in Fig.~\ref{image-train-cnn}.
The used architecture for a convolutional neural network U-Net shown in Fig.~\ref{image-model-unet}.
Train history graph for a convolutional neural network U-Net shown in Fig.~\ref{image-train-unet}.

As a loss function, the mean squared error (MSE) used:
\begin{equation}
    {\rm{MSE}} = (\frac{1}{n})\sum_{i=1}^{n}({Y_i} - \widehat {{Y_i}})^{2}
\end{equation}

The total count of parameters for this neural network models was 28,953,355 parameters for fully convolutional neural network and 29,146,251 parameters for U-Net architecture.
Evaluation of trained models on test data shown in Table \ref{table-loss-cnn} for fully convolutional neural network architecture and in Table \ref{table-loss-unet} for the U-Net model architecture.

\section{Results} \label{Results}

As a result of the training and validation of neural networks, results of the average distance between the largest value of the normal distribution of the mask and the actual coordinate of the anatomical reference point in centimeters obtained.
The final averaged value of the localization error of the anatomical reference points calculated by the cross-validation method.
To evaluate the results obtained, an additional comparison made of the marking results of anatomical reference points between the three doctors participating in the study.
The results of comparing the average distance of localization of anatomical reference points between neural networks and one randomly selected doctor, as well as between three doctors, each with each, are presented in Table \ref{table-compare}.
The average distance values obtained at the cross-validation stages presented in Table \ref{table-cnn} for a fully convolutional neural network and in Table \ref{table-unet} for a convolutional neural network U-Net.
Marking 27 reference points of anatomical references after taking the highest value of the resulting maps for each type of point shown in Fig.~\ref{image-result}

\begin{figure}[ht!]
    \centering
    \centerline{\includegraphics[scale=0.40]{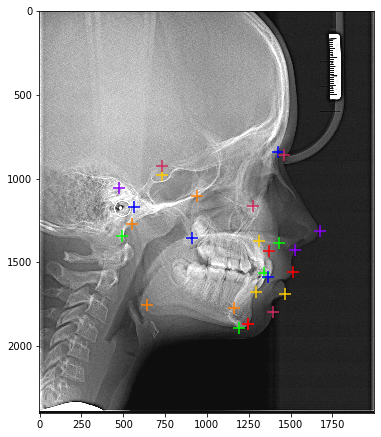}}
    \caption{Result of detecting anatomical reference points.}
    \label{image-result}
\end{figure}

The fully convolutional neural network with extended architecture U-Net has shown results that, in most cases, are better to localization results in a fully convolutional neural network. In 12 of 27 cases, U-Net neural networks showed a result that loses in the accuracy of localization of anatomical reference points by doctors. However, the loss in accuracy of localization in most cases did not exceed 0.5--1.5 cm.

\newpage

\begin{figure}[htbp]
    \centerline{\includegraphics[scale=0.2815]{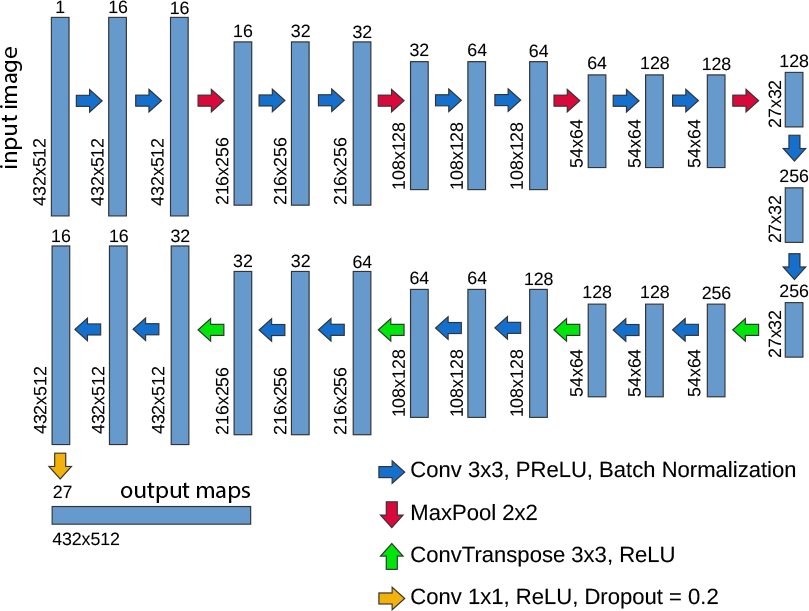}}
    \caption{CNN architecture for image segmentation.}
    \label{image-model-cnn}
\end{figure}

\begin{figure}[htbp]
    \centerline{\includegraphics[width=\columnwidth]{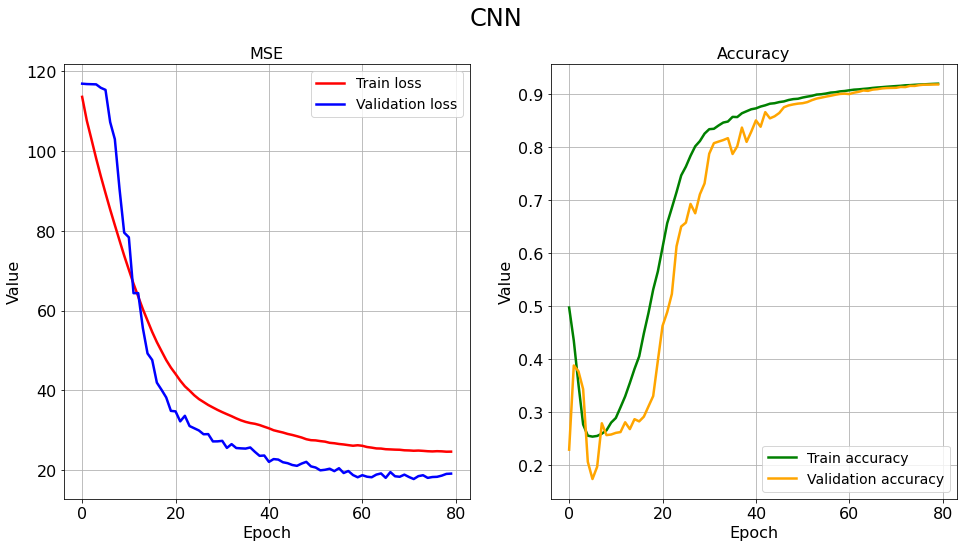}}
    \caption{CNN train history graph.}
    \label{image-train-cnn}
\end{figure}

\begin{table}[htbp!]
    \centering
    \caption{Mean metrics of 5 data splits for CNN with best weights}
    \begin{tabular}{|c|c|c|}
        \hline
        \textbf{Split number} & \textbf{Loss} & \textbf{Accuracy} \\
        \hline
        1                     & 17.809        & 0.914             \\
        2                     & 19.650        & 0.918             \\
        3                     & 19.244        & 0.920             \\
        4                     & 17.408        & 0.922             \\
        5                     & 15.083        & 0.923             \\
        \hline
        \textbf{Mean}         & 17.839        & 0.919             \\
        \hline
    \end{tabular}
    \label{table-loss-cnn}
\end{table}

\newpage

\begin{figure}[htbp]
    \centerline{\includegraphics[scale=0.25]{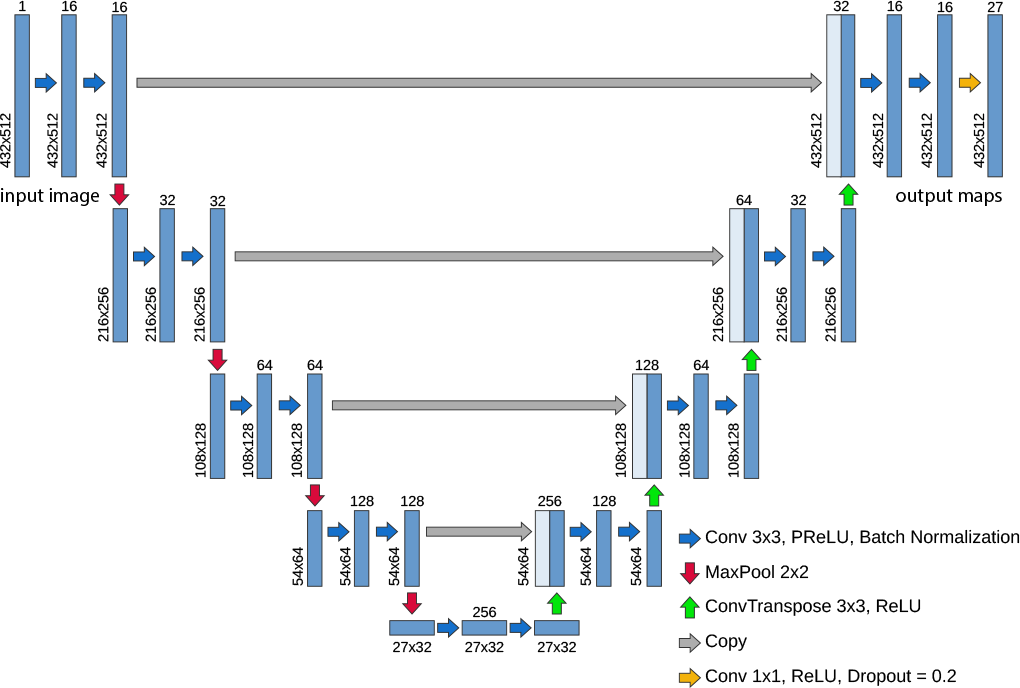}}
    \caption{U-Net architecture for image segmentation.}
    \label{image-model-unet}
\end{figure}

\begin{figure}[htbp]
    \centerline{\includegraphics[width=\columnwidth]{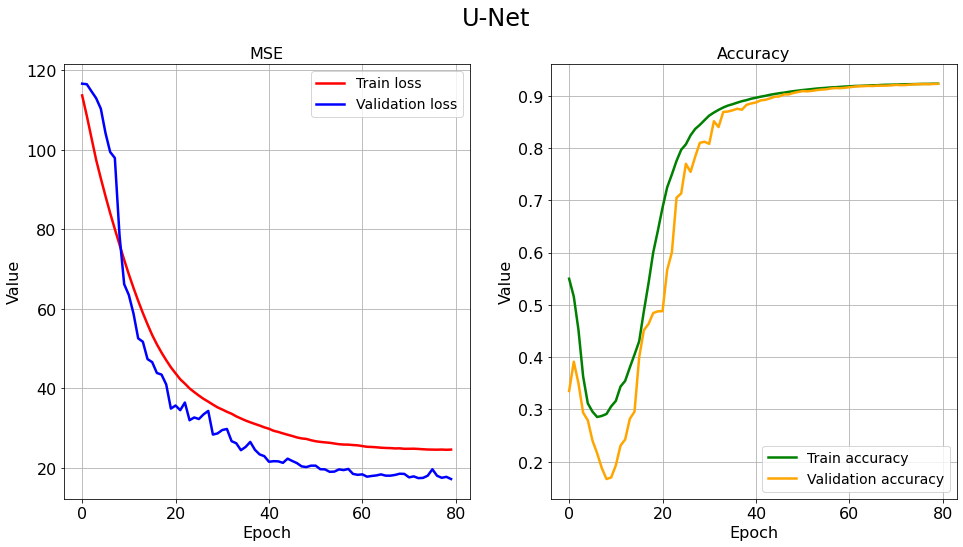}}
    \caption{U-Net train history graph.}
    \label{image-train-unet}
\end{figure}

\begin{table}[htbp!]
    \centering
    \caption{Mean metrics of 5 data splits for U-Net with best weights}
    \begin{tabular}{|c|c|c|}
        \hline
        \textbf{Split number} & \textbf{Loss} & \textbf{Accuracy} \\
        \hline
        1                     & 17.358        & 0.922             \\
        2                     & 19.623        & 0.918             \\
        3                     & 19.536        & 0.921             \\
        4                     & 18.032        & 0.922             \\
        5                     & 15.880        & 0.921             \\
        \hline
        \textbf{Mean}         & 18.086        & 0.921             \\
        \hline
    \end{tabular}
    \label{table-loss-unet}
\end{table}

\begin{table*}[hpt!]
    \caption{Comparison of localization results for CNN, U-Net and doctors}
    \begin{center}
        \begin{tabular}{|c|c|c|c|}
            \hline
            \textbf{Reference} & \multicolumn{3}{|c|}{\textbf{Mean distance, cm}}                                                      \\
            \cline{2-4}
            \textbf{type}      & \textbf{CNN and doctor}                          & \textbf{U-Net and doctor} & \textbf{Three doctors} \\
            \hline
            A                  & 2.26                                             & 1.69                      & 2.15                   \\
            Ar                 & 2.75                                             & 1.85                      & 1.87                   \\
            B                  & 3.52                                             & 3.43                      & 4.08                   \\
            Ba                 & 2.31                                             & 2.99                      & 4.83                   \\
            C                  & 3.42                                             & 3.45                      & 2.98                   \\
            DT pog             & 2.08                                             & 1.91                      & 4.36                   \\
            EN pn              & 2.75                                             & 1.40                      & 1.53                   \\
            Gn                 & 1.84                                             & 1.32                      & 2.66                   \\
            Go                 & 4.02                                             & 3.07                      & 3.42                   \\
            LL                 & 2.30                                             & 1.27                      & 2.22                   \\
            Me                 & 1.85                                             & 1.40                      & 1.67                   \\
            N                  & 1.93                                             & 1.68                      & 0.95                   \\
            Or                 & 4.18                                             & 3.09                      & 4.29                   \\
            Po                 & 3.16                                             & 3.72                      & 9.87                   \\
            Pog                & 2.13                                             & 1.40                      & 4.45                   \\
            Pt                 & 2.75                                             & 2.64                      & 3.59                   \\
            S                  & 2.14                                             & 1.25                      & 0.68                   \\
            SNA                & 2.74                                             & 2.57                      & 1.65                   \\
            SNP pm             & 1.76                                             & 1.46                      & 1.44                   \\
            Se                 & 1.73                                             & 1.65                      & 1.20                   \\
            Sn                 & 1.67                                             & 1.55                      & 0.61                   \\
            UL                 & 2.18                                             & 1.56                      & 1.11                   \\
            aii                & 3.14                                             & 3.03                      & 2.80                   \\
            ais                & 2.88                                             & 2.61                      & 2.45                   \\
            ii                 & 1.76                                             & 1.79                      & 0.71                   \\
            is                 & 1.77                                             & 1.14                      & 0.48                   \\
            n                  & 2.45                                             & 1.93                      & 1.93                   \\
            \hline
            \textbf{Mean}      & 2.50                                             & 2.11                      & 2.59                   \\
            \hline
        \end{tabular}
        \label{table-compare}
    \end{center}
\end{table*}
\begin{table}[hbt!]
    \caption{Distance between anatomical reference true position and predicted position for CNN}
    \begin{center}
        \begin{tabular}{|c|c|c|c|c|c|c|}
            \hline
            \textbf{Reference} & \multicolumn{6}{|c|}{\textbf{Distance, cm}}                                                                                             \\
            \cline{2-7}
            \textbf{type}      & \textbf{Split 1}                            & \textbf{Split 2} & \textbf{Split 3} & \textbf{Split 4} & \textbf{Split 5} & \textbf{Mean} \\
            \hline
            A                  & 2.14                                        & 2.71             & 1.98             & 1.86             & 2.63             & 2.26          \\
            Ar                 & 2.81                                        & 3.10             & 2.99             & 2.29             & 2.57             & 2.75          \\
            B                  & 2.50                                        & 4.91             & 4.09             & 2.74             & 3.33             & 3.52          \\
            Ba                 & 2.62                                        & 2.13             & 2.37             & 2.53             & 1.90             & 2.31          \\
            C                  & 4.17                                        & 4.08             & 2.63             & 3.28             & 2.95             & 3.42          \\
            DT pog             & 1.63                                        & 2.20             & 3.04             & 1.89             & 1.65             & 2.08          \\
            EN pn              & 1.63                                        & 3.13             & 1.27             & 4.78             & 2.92             & 2.75          \\
            Gn                 & 1.80                                        & 1.52             & 1.51             & 2.99             & 1.39             & 1.84          \\
            Go                 & 3.06                                        & 3.32             & 3.75             & 6.79             & 3.18             & 4.02          \\
            LL                 & 1.96                                        & 2.10             & 1.03             & 4.54             & 1.90             & 2.30          \\
            Me                 & 1.63                                        & 1.46             & 1.48             & 1.45             & 3.25             & 1.85          \\
            N                  & 1.66                                        & 2.27             & 2.15             & 1.80             & 1.77             & 1.93          \\
            Or                 & 3.66                                        & 4.51             & 5.78             & 3.44             & 3.49             & 4.18          \\
            Po                 & 3.06                                        & 3.72             & 4.36             & 2.07             & 2.59             & 3.16          \\
            Pog                & 2.14                                        & 1.58             & 2.96             & 2.40             & 1.58             & 2.13          \\
            Pt                 & 2.32                                        & 3.26             & 3.57             & 2.53             & 2.07             & 2.75          \\
            S                  & 1.28                                        & 1.35             & 2.09             & 2.08             & 3.91             & 2.14          \\
            SNA                & 2.93                                        & 2.45             & 3.17             & 2.61             & 2.53             & 2.74          \\
            SNP pm             & 1.95                                        & 1.80             & 1.49             & 1.29             & 2.27             & 1.76          \\
            Se                 & 1.80                                        & 1.94             & 1.94             & 1.69             & 1.28             & 1.73          \\
            Sn                 & 1.58                                        & 1.10             & 2.74             & 1.73             & 1.20             & 1.67          \\
            UL                 & 1.66                                        & 1.99             & 2.95             & 1.14             & 3.17             & 2.18          \\
            aii                & 3.26                                        & 2.84             & 3.23             & 3.96             & 2.41             & 3.14          \\
            ais                & 2.81                                        & 3.28             & 2.67             & 3.22             & 2.42             & 2.88          \\
            ii                 & 1.55                                        & 1.80             & 2.35             & 1.79             & 1.30             & 1.76          \\
            is                 & 1.42                                        & 2.39             & 1.84             & 1.27             & 1.92             & 1.77          \\
            n                  & 2.30                                        & 2.12             & 3.51             & 2.11             & 2.20             & 2.45          \\
            \hline
        \end{tabular}
        \label{table-cnn}
    \end{center}
\end{table}
\begin{table}[hbt!]
    \caption{Distance between anatomical reference true position and predicted position for U-Net}
    \begin{center}
        \begin{tabular}{|c|c|c|c|c|c|c|}
            \hline
            \textbf{Reference} & \multicolumn{6}{|c|}{\textbf{Distance, cm}}                                                                                             \\
            \cline{2-7}
            \textbf{type}      & \textbf{Split 1}                            & \textbf{Split 2} & \textbf{Split 3} & \textbf{Split 4} & \textbf{Split 5} & \textbf{Mean} \\
            \hline
            A                  & 1.56                                        & 1.64             & 1.73             & 1.72             & 1.83             & 1.69          \\
            Ar                 & 1.79                                        & 2.12             & 1.40             & 2.38             & 1.57             & 1.85          \\
            B                  & 2.11                                        & 4.58             & 3.96             & 2.99             & 3.55             & 3.43          \\
            Ba                 & 2.52                                        & 2.05             & 2.38             & 6.27             & 1.74             & 2.99          \\
            C                  & 3.10                                        & 4.16             & 2.66             & 4.02             & 3.30             & 3.45          \\
            DT pog             & 1.53                                        & 1.85             & 3.12             & 1.51             & 1.52             & 1.91          \\
            EN pn              & 1.56                                        & 1.20             & 0.99             & 2.32             & 0.93             & 1.40          \\
            Gn                 & 1.80                                        & 1.44             & 0.92             & 1.11             & 1.32             & 1.32          \\
            Go                 & 2.83                                        & 2.99             & 2.57             & 3.53             & 3.44             & 3.07          \\
            LL                 & 1.48                                        & 1.05             & 0.97             & 1.13             & 1.73             & 1.27          \\
            Me                 & 1.45                                        & 1.49             & 1.36             & 1.51             & 1.20             & 1.40          \\
            N                  & 1.77                                        & 2.41             & 1.24             & 1.52             & 1.46             & 1.68          \\
            Or                 & 3.13                                        & 3.18             & 3.55             & 2.24             & 3.38             & 3.09          \\
            Po                 & 3.82                                        & 3.47             & 4.54             & 3.97             & 2.77             & 3.72          \\
            Pog                & 1.96                                        & 1.49             & 0.85             & 1.03             & 1.68             & 1.40          \\
            Pt                 & 2.45                                        & 2.97             & 3.44             & 2.23             & 2.12             & 2.64          \\
            S                  & 1.00                                        & 1.27             & 1.75             & 1.19             & 1.05             & 1.25          \\
            SNA                & 2.66                                        & 2.16             & 2.84             & 2.85             & 2.33             & 2.57          \\
            SNP pm             & 1.74                                        & 1.72             & 1.28             & 1.30             & 1.26             & 1.46          \\
            Se                 & 1.52                                        & 2.04             & 1.77             & 1.67             & 1.26             & 1.65          \\
            Sn                 & 1.47                                        & 1.06             & 1.82             & 2.18             & 1.25             & 1.55          \\
            UL                 & 1.31                                        & 1.00             & 0.88             & 1.39             & 3.22             & 1.56          \\
            aii                & 2.50                                        & 2.77             & 2.99             & 4.12             & 2.77             & 3.03          \\
            ais                & 2.63                                        & 2.70             & 2.87             & 2.68             & 2.17             & 2.61          \\
            ii                 & 1.36                                        & 1.86             & 2.54             & 1.56             & 1.61             & 1.79          \\
            is                 & 1.18                                        & 0.91             & 1.47             & 1.16             & 0.99             & 1.14          \\
            n                  & 1.57                                        & 2.65             & 1.48             & 2.12             & 1.81             & 1.93          \\
            \hline
        \end{tabular}
        \label{table-unet}
    \end{center}
\end{table}

\newpage

\section{Summary} \label{Summary}

In this paper, we study the effectiveness of deep learning algorithms in solving the problem of detecting anatomical reference points on a radiological image of the head in lateral projection. The process of localizing anatomical reference points described. Effective methods for representing the positions of reference points for the application of deep learning algorithms proposed. As deep learning models, the article examined a fully convolutional neural network and a fully convolutional neural network with an expanded architecture for segmenting biomedical images of U-Net.
Based on the obtained results of the localization of anatomical reference points, we can conclude that the neural networks of the selected architectures can effectively solve the problem.

\section*{Acknowledgment}

The authors gratefully acknowledge the contributions of D. Subbotin and M. Ageeva (FSBEI HE SamSMU MOH Russia) for participating in the preparation of data for this study.

\vspace{12pt}

\end{document}